\documentstyle[11pt]{article}
\title{2+1 Dimensional QED and a Novel Phase Transition}
\author{Thomas Appelquist\\
Department of Physics, Yale University, New Haven, CT 06511
\\ \\
John Terning\\
Department of Physics, Boston University\\
590 Commonwealth Ave., Boston MA  02215
\\  \\
L.C.R. Wijewardhana \\
Department of Physics, University of Cincinnati, Cincinnati,
OH 45221}
\date{February 18, 1994}
\begin{document}
\setlength{\baselineskip}{24pt}
\maketitle
\begin{picture}(0,0)(0,0)
\put(295,310){YCTP-P3-94}
\put(295,299){BUHEP-4-94}
\put(295,288){UC-3-94}
\end{picture}
\vspace{-36pt}

\begin{abstract}
We investigate the chiral phase transition in
2+1 dimensional QED.  Previous gap
equation and lattice Monte-Carlo studies of symmetry breaking have found  that
symmetry breaking ceases to occur when the number of fermion flavors exceeds a
critical value.
Here we focus on the order of the transition.  We find that
there are no light scalar degrees of freedom present  as the critical
number of flavors is approached from above (in the
symmetric phase).  Thus the  phase transition is  not second order, rendering
irrelevant the renormalization group arguments for a fluctuation
induced transition.  However, the order parameter vanishes continuously in the
broken phase,
so this transition is also unlike a conventional first order phase transition.
\end{abstract}

            The study of dynamical mass generation in 2+1 dimensional
gauge theories has attracted a good deal of attention recently.
There are a number of reasons for this interest.
These theories can describe the high temperature thermodynamics
of 4-dimensional relativistic systems, as well as  the
statistical mechanics of certain planar condensed matter
systems. Also, they  have enough structure to be non-trivial,
yet are sufficiently simple to admit analytic solutions, often unlike  their
4-dimensional counterparts. Thus they can serve as theoretical
laboratories for investigating aspects of
dynamical symmetry breaking.

     Mass terms for fermion and gauge fields in 2+1 dimensions
serve as order parameters for discrete as well as
continuous symmetry breaking. Dirac mass terms for 2-component
complex fermions violate parity (P) and time reversal symmetry (T) in 2+1
dimensions. Also, gauge fields admit P and T violating Chern-Simons
mass terms \cite{DJT}. Combining two 2-component fermion fields
it is possible to write down a parity invariant fermion mass term
resembling a 4-dimensional
Dirac mass. If this mass is  dynamically generated, the continuous flavor
symmetry of the system is spontaneously broken from $U(2)$ to
$Sp(2)$.

      The dynamical generation of such a
parity invariant mass term in 2+1 dimensional QED
was investigated by Pisarski \cite{Pisarskia}  using the Schwinger-Dyson (SD)
gap equation. With N four-component Dirac spinors, he used a $1/N$
expansion with
\begin{equation}
\alpha = {{e^2 N} \over{8}}
\label{alpha}
\end{equation}
fixed. Here, $e$ is the
dimensionful gauge coupling. If the fermions
are massless, this theory possesses a $U(2N)$ global flavor symmetry.
If all the four-component fermions acquire the same mass,
then the mass term by itself has an $Sp(2N)$ symmetry.
Pisarski's analysis led him to the conclusion that
there is dynamical mass generation for large N.

A more refined analysis of the gap equation for QED3 \cite{ANW,N} using the
$1/N$ expansion concluded that there is a phase transition at
a  critical number of fermion flavors  $N_c$ ($3 < N_c < 4$) below which
dynamical
mass generation takes place but
above which the fermions remain massless.
This can be understood as follows.
To leading order in the large $N$ limit, the amplitude for the  interaction of
two fermions by the
exchange of a photon is given by $ e^2/k^2(1 + \alpha/k)$. This can
be thought of as a photon propagator of the form $1/k$ multiplied by the
effective dimensionless coupling $e^2/(k + \alpha)$. In the
infrared limit, $k\ll \alpha$ the
effective coupling approaches the infrared
fixed point $8/N$ \cite{appis}.  This infrared coupling weakens like $1/N$ as
$N$ increases due to the screening effect of the fermions.  The dimensionful
parameter $\alpha$ plays only  the role of an ultraviolet
cutoff for this infrared theory.  For $N$ greater than $N_c$, it was then shown
that the effective infrared
coupling is too weak to cause
fermion condensation \cite{ANW}. This is analogous to what happens in 3+1
dimensional gauge theories where the gauge coupling must exceed a critical
value for chiral symmetry breaking to occur.

For $N > N_c$
this analysis \cite{ANW} found nonzero solutions of the gap equation (dynamical
masses) which scale as
\begin{equation}
\Sigma_n(p\approx 0)=\alpha e^{\delta +2} \exp\left({{-2n \pi}\over{\
\sqrt{N_c/N -1}}}\right),
\label{sig0}
\end{equation}
in the limit $N\rightarrow N_c$.  Here, $n=1$,2,3,..., and $\delta$ is a
function of $N$ that is non-singular as
$N\rightarrow N_c$.  It was shown that the $n=1$ solution minimizes the
effective potential, and hence is the ground state solution.
Subsequent lattice
Monte-Carlo analysis \cite{MC} supported this critical behavior in QED3.
Also, estimates of the higher order terms in the $1/N$ expansion
indicated that $N_c$ receives a correction
of no more than $25\%$ \cite{N}.

The gap equation analysis relies on the fact that the effective infrared
coupling is proportional to $1/N$ up to small constant corrections. The
analysis could break down if, in higher orders in the $1/N$ expansion, the
effective coupling received corrections
proportional to powers of $\ln(k/\alpha)$ for $k \ll \alpha$ \cite{Penn}.
It has been shown,
however, that this does not happen to any order in the $1/N$
expansion \cite{AN}, that is, that the infrared fixed point persists to all
orders and the effective infrared theory is scale invariant to all orders.

Recently the question of a phase transition at $N_c$ was re-analyzed by
Pisarski \cite{Pisarskib}
using renormalization group (RG) methods.
He constructed a renormalizable
effective Lagrangian for the
scalar field order parameter for flavor symmetry breaking,
respecting the original $U(2N)$ symmetry of the gauge theory. It was written in
terms of a set of fields $\phi$ transforming as an $SU(2N)$ adjoint.
He neglected the coupling of this field to the fermions, and hence to the gauge
field.
He computed the RG flows of the effective
scalar self couplings and argued that,  when the
quadratic term in the scalar potential is tuned to zero, radiative corrections
generate a non-zero vacuum expectation value for the scalar
field provided $N$ is greater than $\sqrt{5}/2$. He pointed out that this
symmetry breaking is analogous to the
Coleman-Weinberg mechanism
in the Abelian Higgs model \cite{CW} in 3+1 dimensions.
He then invoked the universality hypothesis of Wilson to conclude that
in the original gauge theory, flavor symmetry breaking accompanied by
fermion mass generation will occur for an arbitrarily large number of fermion
flavors.
This conclusion contradicts the gap equation analysis and the lattice
Monte-Carlo results.

The problem with this RG analysis is that universality only relates theories
with identical massless degrees of freedom.
In particular, we will argue that for $N>N_c$ (the symmetric phase),
there are no light scalar degrees of freedom.   For this purpose
one could study the effective potential of this theory, but
this would yield direct information only about the zero-momentum mass, not the
physical
mass of the relevant scalar field. We will use an alternative approach based on
the fact that light composite scalars would show up as poles in
the fermion-antifermion scattering amplitude.  This is the classic method
employed by Nambu and Jona-Lasinio \cite{NJL}.  We will now proceed to solve
the
SD equation  for this scattering amplitude in the symmetric
phase.  We restrict attention to the ladder approximation which was
used in the solution to the gap equation. This approximation should
be as reliable here as it was in the case of the gap equation \cite{N}.  We set
the (Euclidian) momentum of the initial fermion and antifermion to $q/2$, but
keep a non-zero momentum transfer by assigning momenta $q/2 \pm p$ for the
final fermion
and antifermion. If the theory contains a light scalar resonance, the
scattering amplitude should display a pole in (Minkowsky) $q^2$.

If the Dirac indices of the initial fermion and antifermion are $\lambda$ and
$\rho$, and the those of the final state fermion and antifermion are $\sigma$
and $\tau$, then the scattering
amplitude can be written as:
\begin{equation}
T_{\lambda \rho \sigma \tau}(p,q) = \delta_{\lambda \rho} \delta_{\sigma \tau}
T(p,q) + . . .~,
\label{fullscatt}
\end{equation}
where the ... indicates pseudoscalar, vector, axial-vector, and tensor
components.   We contract
Dirac indices so that we obtain the SD equation for the the scalar
s-channel scattering amplitude, $T(p,q)$, containing only t-channel photon
exchanges.  If $p \gg q$, then $q$ will simply act as an infrared cutoff in the
loop integrations.
The SD equation \cite{NN} in the scalar channel
is:
\begin{equation}
T(p,q) = {{ 16\alpha}\over{3 N p^2(1+{\alpha\over p})}}+
{{16 \alpha} \over{3\pi^2 N p}}
\int_q^\infty {{dk \,T(k,q)}\over{k}} \ln\left({{k+p+\alpha}\over{|k-p|+
\alpha}}\right) ~.
\label{T1}
\end{equation}
Note that the first term in equation (\ref{T1}) is simply one photon exchange
in the large
$N$ limit.  It is this large $N$ propagator that is used as the kernel in
deriving  equation (\ref{T1}).
The integral in equation~(\ref{T1}) is rapidly damped for $k > \alpha$.
For $p\ll\alpha$ we use
the approximation:
\begin{equation}
T(p,q) = {{ 16}\over{3 N p}}+{{16} \over{3\pi^2 N p}}
\int_q^\alpha {{dk \,T(k,q)}\over{k}} (k+p-|k-p|) ~.
\label{T2}
\end{equation}

For momenta $p>q$, equation~(\ref{T2}) can be converted to a differential
equation:
\begin{equation}
p{d^2 \over dp^2}\left(p T \right) =
{{-32 \,\,T}\over {3\pi^2 N}}~,
\label{diff}
\end{equation}
with appropriate boundary conditions determined from equation~(\ref{T2}).
The solutions of equation~(\ref{diff}) have the form.
\begin{equation}
T(p,q) = {{A(q)}\over \alpha} \left({p \over \alpha}\right)^{-{1 \over 2} + {1
\over 2} \eta}
 +{{B(q)}\over \alpha} \left({p \over \alpha}\right)^{-{1 \over 2} - {1 \over
2} \eta} ~,
\label{sol}
\end{equation}
where
\begin{equation}
\eta= \sqrt{1-N_c/N}~,
\label{eta}
\end{equation}
and $N_c \equiv 128/3\pi^2$.
The unknown coefficients, $A$ and $B$, can be
determined by substituting the solution back into equation~(\ref{T2}).  This
gives:
\begin{equation}
A ={{-\left({1\over 2}-{1\over 2}\eta\right)^2 \pi^2  \left({q\over \alpha}
\right)^{-{1\over 2}+{1\over 2}\eta} }
\over {2 \left({1\over 2}+{1\over 2}\eta\right) \left(1-
\left({{1-\eta}\over{1+\eta}}\right)^2 \left({q\over \alpha}
\right)^\eta \right)}   }
{}~,
\label{A}
\end{equation}
and
\begin{equation}
B ={{\left({1\over 2}-{1\over 2}\eta\right) \pi^2  \left({q\over \alpha}
\right)^{-{1\over 2}+{1\over 2}\eta} }
\over {2 \left(1-
\left({{1-\eta}\over{1+\eta}}\right)^2 \left({q\over \alpha}
\right)^\eta \right)}   }~.
\label{B}
\end{equation}

Note that there is an infrared divergence in the limit $q \rightarrow 0$ in the
numerators of both
(\ref{A}) and (\ref{B}).  That this is an infrared divergence rather
than a pole corresponding to a bound state can be seen from the fact that
the divergence exists for arbitrarily weak coupling ($1/N$).  In fact,
this infrared divergence can be seen at order $1/N^2$ in the one-loop (two
photon exchange) diagram.

If we denote the location of the poles of the functions $A$ and $B$ in the
complex $q$ plane by $q_0$, we
then have
\begin{equation}
|q_0| = \alpha \left({{1+\eta}\over{1-\eta}}\right)^{2 \over \eta}~.
\label{q0}
\end{equation}
In the limit that $\eta \rightarrow 0$, we have
\begin{equation}
|q_0| \rightarrow \alpha \exp\left(4\right) ~.
\label{q0limit}
\end{equation}
Thus we see there are no poles in the complex $q_0$-plane within the domain of
validity of our
approximations (i.e. $|q_0| \ll \alpha$), as $\eta\rightarrow 0$. In particular
there is no pole which approaches zero momentum as $\eta\rightarrow0$.  This is
the main result of
this letter. The non-appearance of a scalar whose physical mass approaches zero
at the the critical coupling \cite{Miransky}, $N_c$, demonstrates that the
phase transition is
not second order.

How does this result match on to the gap-equation analysis
in the broken phase and the lattice Monte-Carlo calculations?
Our result may seem to contradict these previous studies, since
equation (\ref{sig0}) shows that the dynamical mass $\Sigma_n$ (which can
serve as an order parameter for the chiral phase transition) vanishes
continuously as $N \rightarrow N_c$ in the broken phase ($N<N_c$).
Note, however,  that the order parameter, $\phi$, of a transition with a finite
correlation length (i.e. {\em not} second order) may
vanish continuously if the effective potential is not analytic at $\phi=0$.
In quantum field theories with long range forces,
effective potentials are generally not analytic at $\phi=0$.

Recall that there are an infinite number of solutions (see equation
(\ref{sig0})) for the dynamical mass (in the broken phase) labeled by
the integer $n$.  All of these solutions must correspond to extrema of the
effective potential.  Consider the effective potential $V(m)$ as a
function of the dynamical mass, $m$, of the fermion at zero momentum.  Even
without
an explicit calculation of $V(m)$, we can infer some of its properties from the
knowledge we
already have about its extrema.  When
$m=\Sigma_n(p=0)$, the mass is at a local minimum or maximum of $V(m)$.  Since
the global minimum is at $m=\Sigma_1(p=0)$, we can infer that there is a local
maximum at $m=\Sigma_2(p=0)$, and a local minimum at $m=\Sigma_3(p=0)$, and so
on.  Note that $\Sigma_1(p=0)>\Sigma_2(p=0)>\Sigma_3(p=0)>$ ..., so there
are an infinite number of local minima (metastable states) between the
symmetric (false vacuum) state at the origin ($m=0$) and the true ground state
at
$m=\Sigma_1(p=0)$.  Thus there are an infinite number of energy barriers
between the
symmetric state and the true ground state.  This is suggestive of a first order
transition.  However, we note that we can take $\Sigma_1(p=0)$ to be an order
parameter for the phase transition,
and as we approach the critical coupling, $N_c$, the order parameter approaches
zero, and we expect a scalar bound state with a mass of the order of
$2\Sigma_1(p=0)$.
This latter behavior is more typical of a second order transition.  However the
spectrum is not continuous as we go through the critical value  $N_c$, since
there
is no light scalar in the symmetric phase.  This unusual critical behavior can
be attributed to the fact that the effective potential is not analytic at
$m=0$.

To conclude, we have re-examined the critical behavior of QED3 as a function of
the number of fermions,  $N$, by solving a Schwinger-Dyson equation for the
fermion-antifermion scattering amplitude in the symmetric phase ($N>N_c$).\
We have argued that no light scalar degrees of freedom appear in the
symmetric phase, and hence that the chiral symmetry breaking phase transition
is not second order.  Since the order parameter varies continuously
from the broken to the symmetric phase, this does not look like a
conventional first order phase transition.  This unusual
behavior can be attributed to the presence of long range forces \cite{FMN} in
this model.
We have also described the likely structure of the effective
potential in the broken phase, and noted the unusual features of the critical
behavior arising from the non-analyticity of the effective potential.  The fact
that
there are no light scalar degrees of freedom in the symmetric phase indicates
that
Pisarski's analysis \cite{Pisarskib} is not relevant to the chiral phase
transition in
QED3.  Another way to say this is that the $SU(2N)$ scalar field theory he
analyses is not in a
universality class with QED3.  (Indeed, since the chiral phase transition in
QED3 is not second order, QED3 does not have a universality class.)  In fact,
it is not surprising that  QED3 and the $SU(2N)$ scalar theory have different
phase transitions, since universality arguments are
only expected to be applicable to theories with finite-range interactions, a
property
that gauge theories do not enjoy.

 It would be interesting to study chiral phase
transitions in other gauge theories  \cite{ATWNEXT}.   Of  particular interest
in four dimensions are  the zero temperature  QED transition and the finite
temperature QCD transition.

\noindent \medskip\centerline{\bf Acknowledgements}
\vskip 0.15 truein
We would like to thank M. Carena, S. Chivukula, J. Marko, S. Sachdev, H.
Saleur,
R. Shankar, R. Sundrum, and C. Wagner for helpful
discussions.
We also thank the Aspen Center for Physics where part of this work was
completed.
This work was supported in part by the Natural Sciences and
Engineering Research Council of Canada; the Texas National Research
Laboratory Commission  through an SSC fellowship and under contracts
\#RGFY93-278 and  \#RGFY93-272;  and by the Department of Energy under
contracts \#DE-FG02-92ER40704 and \#DE-FG02-91ER40676,  \#DE-FG-02-84ER40153.
\vskip 0.15 truein



\begin{thebibliography}{99}
\bibitem{NN} For convenience, we use the gauge where the fermion
wavefunction renormalization vanishes (i.e. gauge parameter $\xi=2/3$),
see  H. Nakatani and ref. \cite{N}.
\end{thebibliography}
\end{document}